\documentclass[a4paper,12pt]{article}

\usepackage{epsf}
\title{Classical and Quantum Limits in Bohmian Quantum Cosmology}
\author{Fatimah Shojai$^{1,2}$\thanks{fatimah@theory.ipm.ac.ir} \& Abbas Shirinifard$^3$\thanks{shirinifard@mehr.sharif.ir}}
\begin{document}
\maketitle
\begin{center}
$^1$ Department of Physics, University of Tehran,Tehran, Iran;\\
$^2$Institute for Studies in Theoretical Physics and Mathematics, Tehran, Iran;\\
$^3$ Department of Physics, Sharif University of Technology, Tehran,
Iran.
\end{center}
\begin{abstract}
In this paper we have investigated the classical limit in Bohmian quantum cosmology.
It is observed that in the quantum regime where the quantum potential is greater than the
classical one, one has an expansion in terms of negative powers of the Planck constant. But 
in the classical limit there are regions having positive powers of the Planck constant, and
regions having negative powers and also regions having both. The conclusion is that the
Bohmian classical limit cannot be obtained by letting the Planck constant goes to zero.
\end{abstract}
\section{Introduction and survey}
In the semi classical approximation of quantum mechanics the quantum effects are small 
perturbation around classical physics and can be obtained via WKB approximation. WKB has the aim to obtain an asymptotic solution of Schrodinger equation. It is an approximation, because one assums a slowly varing potential. If $K^2=\frac{2m}{\hbar^2}(E-V(x))$ and the charactristic lenght of the system be denoted by $L$, WKB approximation means $KL>>1$ and the wave function is expanded asymptotically in powers of $\frac{1}{KL}$. So in this approximation the quantum correction is expressed in terms positive 
powers of $\hbar$. Assuming WKB conditions, two leading terms are sufficient to have the essential 
part of expansion. The role of higher order terms is investigated in \cite{blas}. 
There are many arguments against this definition of semi classical limit.
 For example it is incompatible with the superposition principle of quantum mechanics
 and it is completely different from Ehrenfest's theorem \cite{bal}.

Moreover the precise definition of classical limit depends on 
one's approach to quantum theory. The best example is  de-Broglie--Bohm theory of 
quantum mechanics \cite{bohm}. In this theory the quantum effects are introduced by adding 
a term, 
\textit{quantum potential}, to the classical potential. It depends on the second  
derivative of the norm of the wave function. So  the classical limit is defined 
according to the 
value of the quantum potential as compared with the classical potential. 
Choosing a special form of the wave function for  which the quantum potential is ignorable, 
represents a semi classical wave function. This is completely different from WKB conditions. The conditions of WKB approximation are not sufficient to 
guarantee that the quantum
 potential is small and  varies slowly as compared to the classical potential\cite{hol}.   
 In de-Broglie--Bohm's viewpoint there are many different works 
on how one can approach to the classical limit \cite{hol}. Arguments on the basis of 
interaction potential and boundary conditions \cite{mos}, introducing the classical 
limit as a scaling limit \cite{allori} and the existence of significant quantum potential for a
 sharp wave function by Ballentine \cite{bal} are some of them. The important fact is that
  it is quiet possible to have a 
precise definition of the classical limit in de-Broglie--Bohm theory without any reference to a perturbation 
expansion. 

In this paper, following de-Broglie--Bohm's description, we shall discuss three regimes
 for a gravitational model via a perturbative expansion\footnote{We consider a gravitational
 model here, but the discussion is quiet general and can be applied to any system.}. 
 Classical regime, which incorporates quantum effects as corrections to the classical 
 Hamilton--Jacobi function, quantum regime which is the opposite of classical regime,
  and the transition region between quantum mechanical and classical behavior. For this,
   firstly, we present a perturbation expansion of de-Broglie--Bohm's  equations in 
terms of positive powers of $\hbar$. As one expects the first order is pure classical. 
 
This motivates us to look for a perturbation expansion of de-Broglie--Bohm's 
equations which is suitable for quantum regime. Those states which belong 
to the quantum regime necessarily have not a semi classical limit. Recently some highly 
non--classical quantum states are 
investigated in ref \cite{foldi}. As the quantum regime stands for the opposite 
of classical regime, it is expected that the classical effects be a small 
perturbation of quantum physics in the quantum realm. Now the question is which 
expansion in terms of $\hbar$ is suitable in this limit? In the next sections we shall 
see 
that the classical corrections must be introduced by negative powers of $\hbar$. We 
see that if one writes a decreasing expansion in terms of negative powers of $\hbar$ 
the pure quantum effects appear at the first order. Moreover in the higher 
orders of expansion some classical effects emerge as a perturbation to the pure 
quantum solution. A similar series of decreasing powers of $\hbar$ introduced by
 Bronzan\cite{B1} as the opposite limit to the WKB approximation. This method firstly
 applied to the energy of weakly bound particle in one dimension and then it 
 is generalized to a modified WKB approximation \cite{B2} which is basically 
 different from our method which emphasizes on an expansion in terms of negative 
 powers
 of $\hbar$ for quantum regime. 

In both cases, the classical and quantum realm, the condition that the higher order 
corrections are  smaller and smaller, leads to a validity domain for the scale 
factor of the universe.   
\section{A MiniSuperSpace}
In order to justify the points mentioned in the previous section, here we shall 
consider the application of de-Broglie--Bohm quantum mechanics to the WDW 
quantum gravity. And to make the argument more simple we consider the 
case of a minisuperspace, FRW universe. 
The WDW equation for such a system is:
\begin{equation}
\hbar^2\frac{d^2\psi(a)}{da^2}-a^2(k-\Lambda a^2)\psi(a)=0
\end{equation}
Here we have chosen $16\pi G/c^4=1$, $k$ is the curvature parameter and 
$\Lambda$ is the cosmological constant.

The Bohmian equations for such a system are given by\footnote{This can be achieved by setting $\psi=R \exp(iS/\hbar)$.}\cite{hor}:
\begin{equation}
\left (\frac{d S}{da}\right )^2+V+Q=0
\label{z2}
\end{equation}
where $S$ is the Hamilton--Jacobi function, $V=a^2(k-\Lambda a^2)$ is the classical
potential and the quantum potential is defined as:
\begin{equation}
Q=-\hbar^2\frac{d^2R/da^2}{R}
\end{equation}
in which $R^2$ gives the ensemble density of the system and thus satisfying
the continuity equation as:
\begin{equation}
\frac{d}{da}\left ( R^2\frac{dS}{da} \right )=0
\end{equation}
This has the solution:
\begin{equation}
R^2\frac{dS}{da}=cons.
\label{z1}
\end{equation}
The Bohmian trajectories are given by the guidance relation:
\begin{equation}
-a\frac{da}{dt}=\frac{dS}{da}
\end{equation}
As it is stated in the previous section, in the viewpoint of de-Broglie--Bohm 
theory, the classical limit is determined by criteria of quantum potential
and quantum force being much less than their classical counterparts and this does not necessarily means the limit $\hbar\rightarrow 0$. In order to 
clarify this we use the above minisuperspace model. Choosing $k=0$, one can 
obtain an exact solution:
\begin{equation}
\psi(a)=\sqrt{a}\left (J_{1/6}\left (\frac{\sqrt{\Lambda}a^3}{3\hbar}\right ) 
+\beta Y_{1/6}\left (\frac{\sqrt{\Lambda}a^3}{3\hbar}\right ) \right )
\end{equation}
Let us investigate this wave function in the limit of large scale factor, and let's choose:
\begin{equation}
\beta=-i+\delta
\end{equation}
where $\delta$ is an infinitesimal real number. Setting $\delta=0$ we get 
the classical limit,
because the phase would be the classical Hamilton--Jacobi function, using the asymptotic expansion
of the Bessel functions:
\begin{equation}
S_{class}=-\frac{\sqrt{\Lambda}a^3}{3}+\frac{\hbar\pi}{3}
\end{equation}
For a small $\delta$ we have:
\begin{equation}
S=S_{class}-\hbar\delta\sin^2(S_{class}/\hbar)-\hbar\delta^2
\sin^4(S_{class}/\hbar)\tan(S_{class}/\hbar)+\cdots
\end{equation}
which clearly has inverse powers of $\hbar$. But the quantum potential has the expansion:
\begin{equation}
Q=-\frac{2\hbar^2}{a^2}+2\Lambda\delta a^4 \sin\left ( 
\frac{\sqrt{\Lambda}a^3}{3}-\frac{\pi}{3} \right )+{\cal O}(\delta^2)
\end{equation}
It is seen that the quantum potential is  small as compared to the classical potential:
\begin{equation}
\left | \frac{Q}{V}\right | \sim 2\delta \sin\left ( 
\frac{\sqrt{\Lambda}a^3}{3}-\frac{\pi}{3} \right ) << 1
\end{equation}
and 
thus the classical limit is obtained but with negative powers of $\hbar$ in
the expansion of Hamilton--Jacobi function.

In the following sections, we use this fact, and expand the Hamilton--Jacobi
function in terms
of increasing and decreasing powers of $\hbar$ and obtain the regime of 
validity 
of these expansions for the above minisuperspace.
\section{Expansion in terms of increasing powers of $\hbar$}
Let us first examine the case of possible solutions with increasing powers of
$\hbar$. Using the relation (\ref{z1}) the Bohmian equation for the Hamilton--Jacobi 
function (\ref{z2}) for the flat case can be written as:
\begin{equation}
X^4+X^2V+\frac{\alpha}{2}\left ( XX''-\frac{3}{2}X'^2\right )=0
\label{z3}
\end{equation}
with $X=\frac{dS}{da}$ and $V=-\Lambda a^4$ and 
$\alpha=\hbar^2$. Suppose that we have expanded X in terms of increasing powers of $\alpha$ as: 
\begin{equation}
X=X_0+\alpha X_1+\alpha^2 X_2 + \cdots
\end{equation} 
Putting this expansion in the equation (\ref{z3}), one shall get the following algebraic relations 
order by order.
\begin{itemize}
\item{\bf Zeroth Order} 

In this order we have:
\begin{equation}
X_0^4+X_0^2V=0
\end{equation}
with the solution:
\begin{equation}
X_0=-\sqrt{-V}
\end{equation}
which is just the classical relation and corresponds to the leading term of WKB expansion.
\item{\bf First Order} 

Having the zeroth order solution, the first order solution is simple to derive. 
The equation (\ref{z3}) leads to:
\begin{equation}
4X_0^3X_1+2X_0X_1V-\frac{3}{4}X^{'2}_0+\frac{1}{2}X_0X''_0=0
\end{equation}
Its solution is:
\begin{equation}
X_1=\frac{3X^{'2}_0-2X_0X''_0}{8X_0(V+2X_0^2)}
\end{equation}

At any order, the relation is an algebraic relation in terms of 
previous orders, so the solution can be easily obtained at each order.
\end{itemize}
Using the relation (\ref{z1}) one can obtain the expansion of $R$ as:
\begin{equation}
R\propto \frac{1}{\sqrt{|X_0|}}-\frac{\alpha X_1}{2|X_0|^{3/2}}-\alpha^2 
\frac{4X_0X_2-3X_1^2}{8|X_0|^{5/2}}+ \cdots
\end{equation}
And this would leads to the following expansion of quantum potential:
\begin{equation}
Q=\alpha \frac{2X''_0X_0-3X_0^{'2}}{4X_0^2}+\alpha^2
\frac{X''_1X_0^2-3X'_1X'_0X_0+3X_1X_0^{'2}-X_1X_0X''_0}{2X_0^3} +\cdots
\end{equation}
The important problem is that the quantum potential is also an expansion in
terms of increasing powers of $\alpha$, so the limit $\hbar\rightarrow 0$ and
$Q\rightarrow 0$ are the same, and thus the classical limit is determined by going
to the limit $\hbar\rightarrow 0$, for the region that the above expansion is correct.

In order to see the validity regime of the expansion in terms of increasing
powers of $\alpha$ let us evaluate $X$ by inserting the value of the classical 
potential. The result is:
\begin{equation}
X_0=-\sqrt{a^2(a^2\Lambda-k)}
\end{equation}
\begin{equation}
X_1=\frac{8a^4\Lambda^2-6ka^2\Lambda+3k^2}{8a^3(a^2\Lambda-2k)(a^2\Lambda-k)^{3/2}}
\end{equation}
and for the quantum potential we have:
\begin{equation}
Q=\frac{-2\alpha}{a^2}+\frac{27\alpha^2}{\Lambda a^8}+\cdots
\end{equation}
It can be simply seen that in the limit of large $a$, the expansion is appropriate,
because
\begin{equation}
\lim_{a\rightarrow\infty} X_0=-a^2\sqrt{\Lambda}
\end{equation}
\begin{equation}
\lim_{a\rightarrow\infty} X_1=\frac{1}{a^4\sqrt{\Lambda}}
\end{equation}
and in this limit $|\alpha X_1|<<|X_0|$, provided that
$a>>a_c$ with
\begin{equation}
a_c=\left (\frac{\alpha}{\Lambda}\right )^{1/6}=(\Lambda\ell_p^2)^{-1/6}\ell_p
\end{equation}
 so one can use this expansion for large scale factors.

\begin{figure}[htb]
\begin{center}
\epsffile{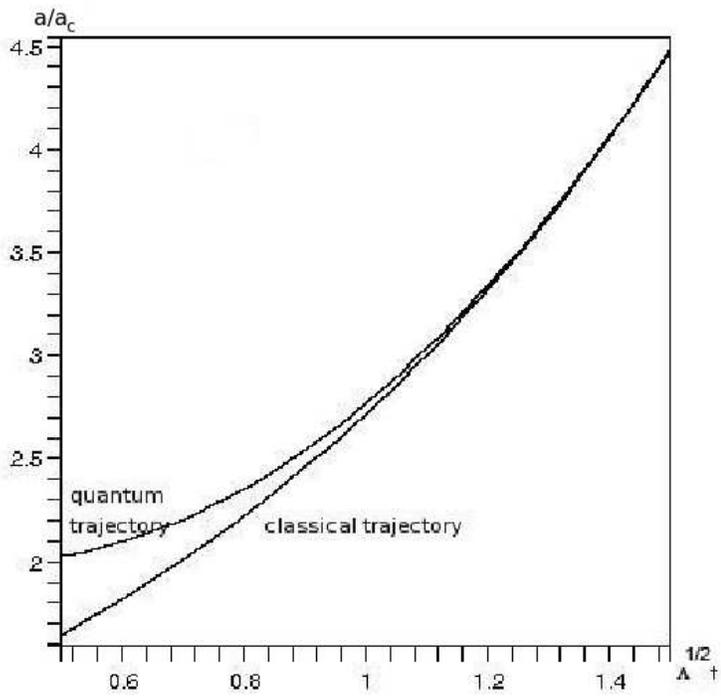}
\end{center}
\caption{Classical and quantum trajectories of the scale factor as 
a function of time, for the regime of large scale factors.}
\label{f1}
\end{figure}

\begin{figure}[htb]
\begin{center}
\epsffile{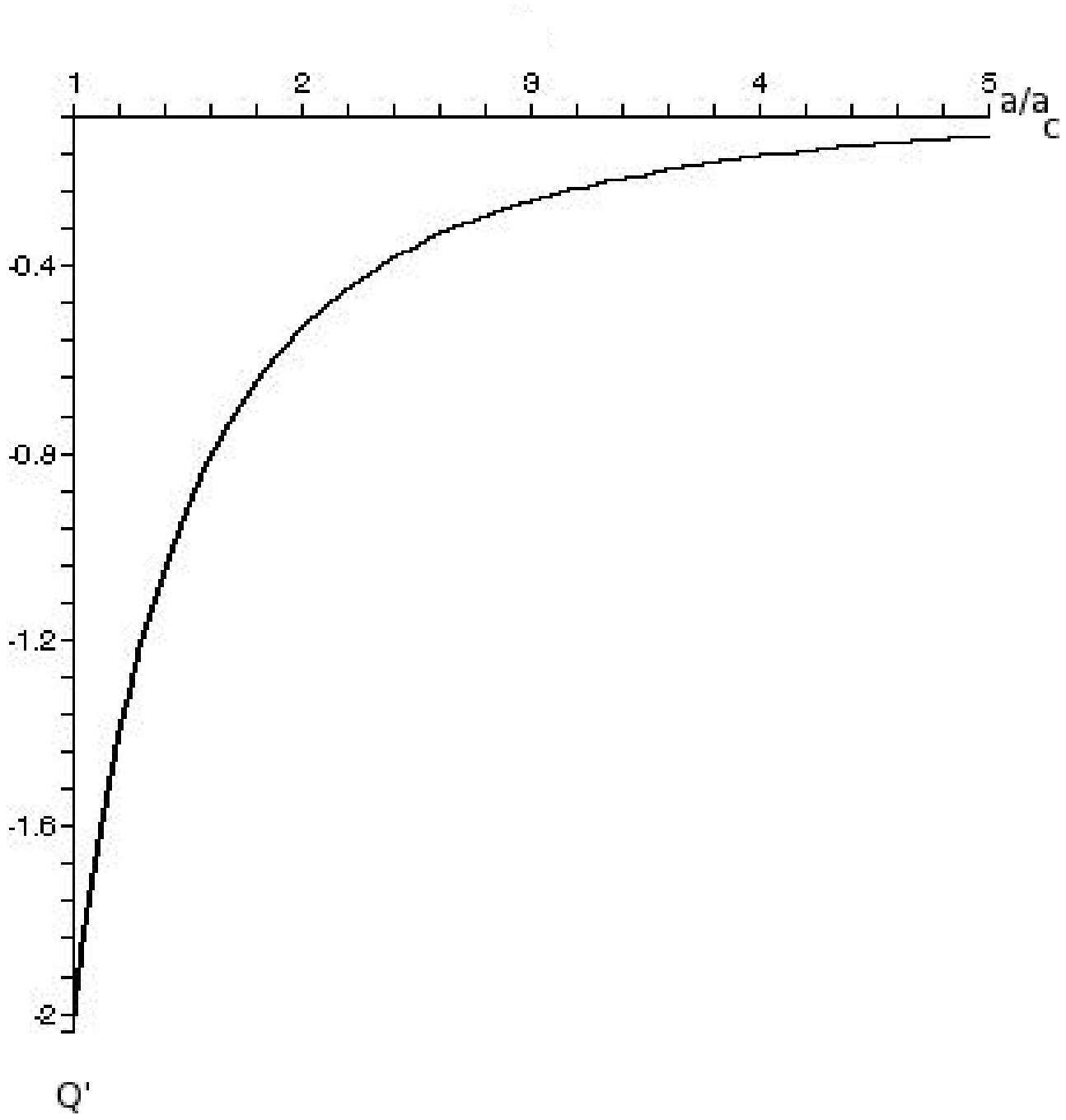}
\end{center}
\caption{Quantum potential ($Q'=\alpha^{-2/3}\Lambda^{-1/3}Q$) as a function of scale factor, for the regime 
of large scale factors.}
\label{f2}
\end{figure}

At this end it is useful to derive the Bohmian trajectories, using the guidance relation:
\begin{equation}
-a\dot{a}=X=X_0+\alpha X_1+\cdots
\end{equation}
Up to zeroth order, we have the classical solution:
\begin{equation}
a=a_0\exp(\sqrt{\Lambda}t)
\end{equation}
and up to the first order:
\begin{equation}
a^6=\frac{\alpha}{\Lambda}+a_0^6\exp(6\sqrt{\Lambda} t)
\end{equation}
The classical and quantum trajectories are plotted in figure (\ref{f1}). Scale factor
is scaled by $a_c$
and time is scaled by $1/\sqrt{\Lambda}$ and a value of $a_0$ is chosen so that the 
difference between the classical and quantum trajectories can be seen easily. As it 
is explained above, the quantum trajectory converges to the classical one at large times.
\epsfxsize=4in \epsfysize=4in
In figure (\ref{f2}), we have plotted the quantum potential (in fact $Q'=\alpha^{-2/3}\Lambda^{-1/3}Q$)
 as a function of the scale
factor, up to the first order. As it is seen, the quantum potential is 
negligible compared to the classical one, therefore we are in the classical limit. 
\epsfxsize=4in \epsfysize=4in
\section{Expansion in terms of decreasing powers of $\hbar$}
In this section we shall study the possibility of having a solution with decreasing
powers of $\hbar$. Again one starts from equation (\ref{z3}) and expand $X$ in 
terms of decreasing powers of $\alpha$. In the domain of applicability of this 
expansion the classical limit is not necessarily present. Therefore the expansion is 
as follows:
\begin{equation}
X=\alpha^{1/2}\left ( X_0+\alpha^{-1}X_1+\alpha^{-2}X_2+\cdots \right )
\end{equation}
Just like the previous section we put this expansion in the equation (\ref{z3}), 
and this time 
we shall get differential equation at each order. (In contrast to the increasing 
power case, where we had algebraic relations.)
\begin{itemize}
\item{\bf Zeroth Order} 

The equation would be:
\begin{equation}
X_0^4+\frac{1}{2}\left (X_0X''_0-\frac{3}{2}X_0^{'2}\right )=0
\end{equation}
with the exact solution:
\begin{equation}
X_0=\frac{4C_1}{16+C_1^2C_2^2+2C_1^2C_2a+C_1^2a^2}
\end{equation}
where $C_1$ and $C_2$ are two constants of integration. Note that there is no footstep
of the classical potential here and this shows that this is just a pure quantum
solution (at this order).
\item{\bf First Order} 

The equation (\ref{z3}) at this order leads to:
\begin{equation}
4X_0^3X_1+X_0^2V+\frac{1}{2}\left ( X''_0X_1+X_0X''_1-3X'_0X'_1\right )=0
\end{equation}
The solution again can be obtained simply as:
\[ 
X_1=\frac{1}{(16+C_1^2(a+C_2)^2)^2}\left (
C_4(a+C_2)+C_3(C_1^2a^2-C_1^2C_2^2-16)\right .
\]
\[
+8C_1(a+C_2)\int da V(a)(C_1^2a^2-C_1^2C_2^2-16)
\]
\begin{equation}
\left . -8C_1\int da V(a)(C_1^2a^2-C_1^2C_2^2-16)(a+C_2) \right )
\end{equation}
\end{itemize}
For simplicity we choose the case $k=0$ and the initial conditions 
$C_2=C_3=C_4=0$. Also we put $C_1=\sqrt{\Lambda}\beta_1$ with $\beta_1$ a dimensionless
constant. 
With these parameters, we have:
\begin{equation}
\alpha^{1/2}X_0=\frac{4\alpha^{1/2}\sqrt{\Lambda}\beta_1}{16+\Lambda\beta_1^2 a^2}
\end{equation}
and
\begin{equation}
\alpha^{-1/2}X_1=\frac{1}{\alpha^{1/2}}
\frac{4\Lambda^{3/2}\beta_1a^6\left (5\Lambda\beta_1^2a^2+112\right )}{105(16+\Lambda\beta_1^2 a^2)^2}
\end{equation}
Again the domain of validity of this expansion is given by 
$|\alpha^{-1}X_1|<<|X_0|$ leading to $a<<a_c$.

\begin{figure}[htb]
\begin{center}
\epsffile{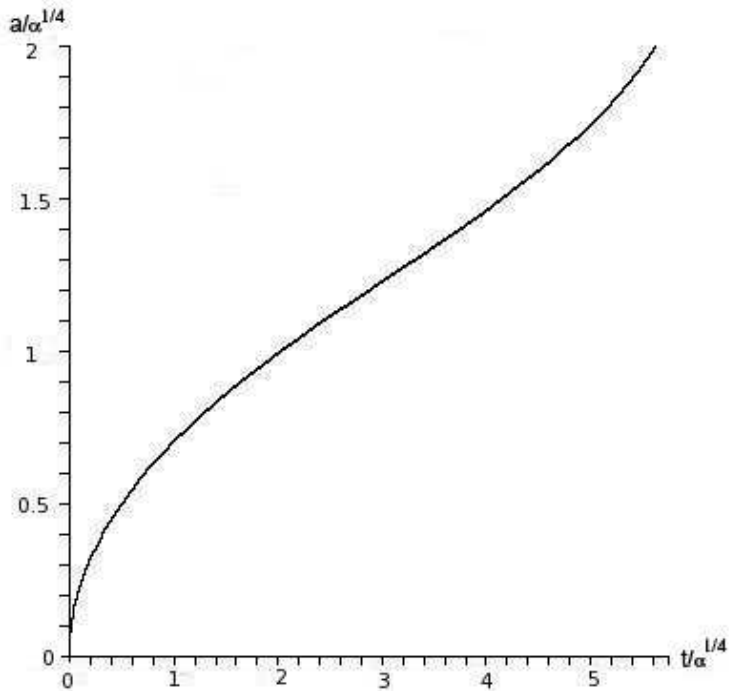}
\end{center}
\caption{Quantum trajectory of the scale factor as 
a function of time, for the regime of small scale factors.}
\label{f3}
\end{figure}

\begin{figure}[htb]
\begin{center}
\epsffile{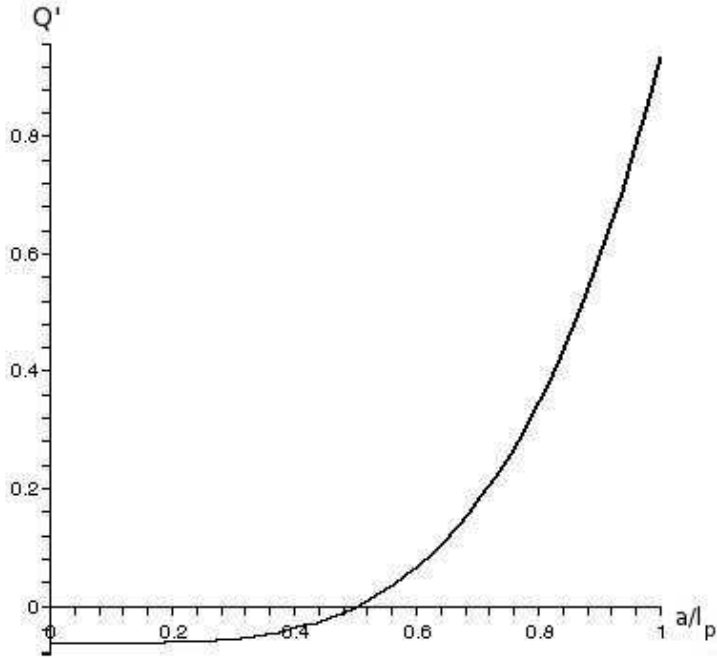}
\end{center}
\caption{Quantum potential ($Q'=Q/\Lambda\alpha$) as a function of scale factor, 
for the regime of small scale factors.}
\label{f4}
\end{figure}

One can obtain an expansion for the norm of the wave function and the quantum potential
using relation (\ref{z1}), as:
\begin{equation}
R=\alpha^{-1/4}\left ( \frac{1}{\sqrt{|X_0|}}-\alpha^{-1}
\frac{X_1}{2|X_0|^{3/2}}+\cdots \right )
\end{equation}
and
\begin{equation}
Q=\alpha \frac{2X''_0X_0-3X_0^{'2}}{4X_0^2}+
\frac{X''_1X_0^2-3X'_1X'_0X_0+3X_1X_0^{'2}-X_1X_0X''_0}{2X_0^3} +{\cal O}(\alpha^{-1})
\end{equation}
With the above values of functions $X_0$ and $X_1$ we have:
\begin{equation}
Q=\frac{-16\alpha\Lambda}{(16+\Lambda a^2)^2}+\Lambda a^4-\frac{32\Lambda a^6\left ( 5\Lambda a^2+112\right )}{105 (16+\Lambda a^2)^3}+\cdots
\end{equation}
The quantum trajectory is plotted in figure (\ref{f3}), 
and as it can be seen the solution is completely different from the classical
solution. In figure (\ref{f4}) we have plotted the quantum potential ($Q'=Q/\Lambda\alpha$) for 
this regime.

It must be noted that this regime ($a<<a_c$) which admits an expansion in terms of negative powers of $\hbar$ can be devided into two intervals. In the first interval, $a_q<<a<<a_c$, the quantum potential is negligible compared to the classical one (classical limit) but in the second, $0<a<<a_q$, the quantum potential is not be negligible (quantum limit). In order to find $a_q$, one should set $|Q|\sim |V|$ leading to $a_q\sim l_p$. 
\epsfxsize=4in \epsfysize=4in
\epsfxsize=4in \epsfysize=4in
\section{Conclusion}
We have shown that since the classical limit in de-Broglie--Bohm theory has
the meaning that the quantum potential is much less than the classical 
potential and the quantum force is much less than the classical force, it is
quite possible to have negative powers of $\hbar$ in the Hamilton--Jacobi
function, in the classical limit.

\begin{figure}[htb]
\begin{center}
\epsffile{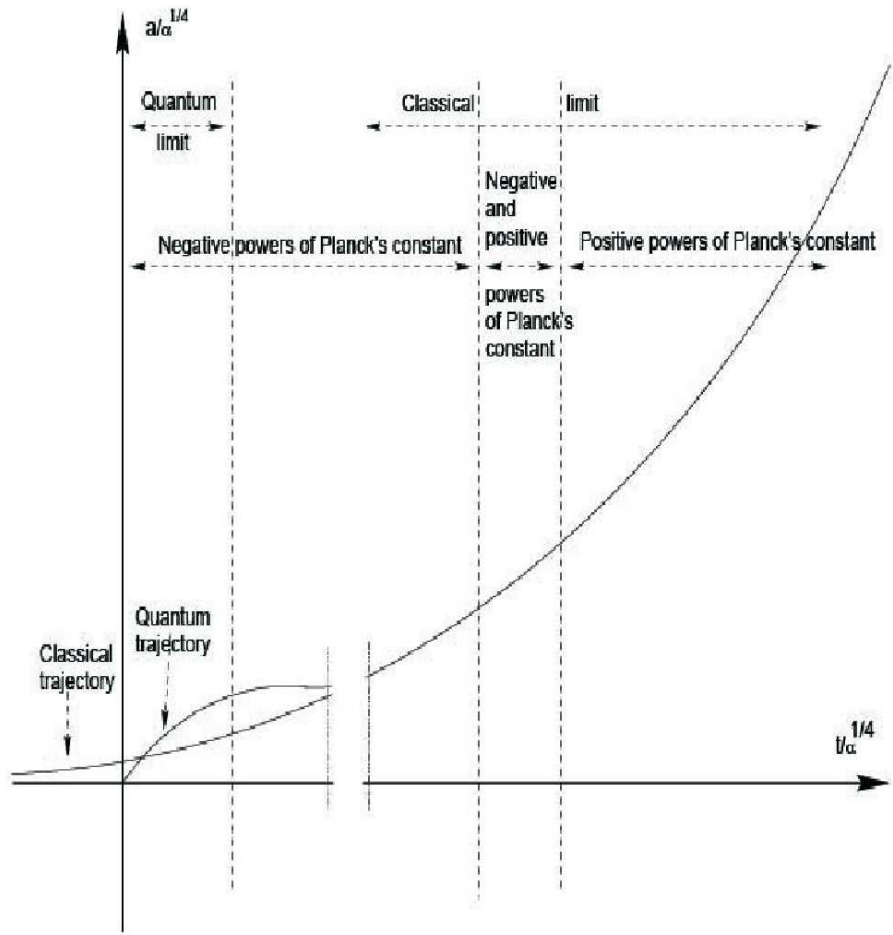}
\end{center}
\caption{Classical and Quantum 
trajectories.}
\label{f5}
\end{figure}

\begin{figure}[htb]
\begin{center}
\epsffile{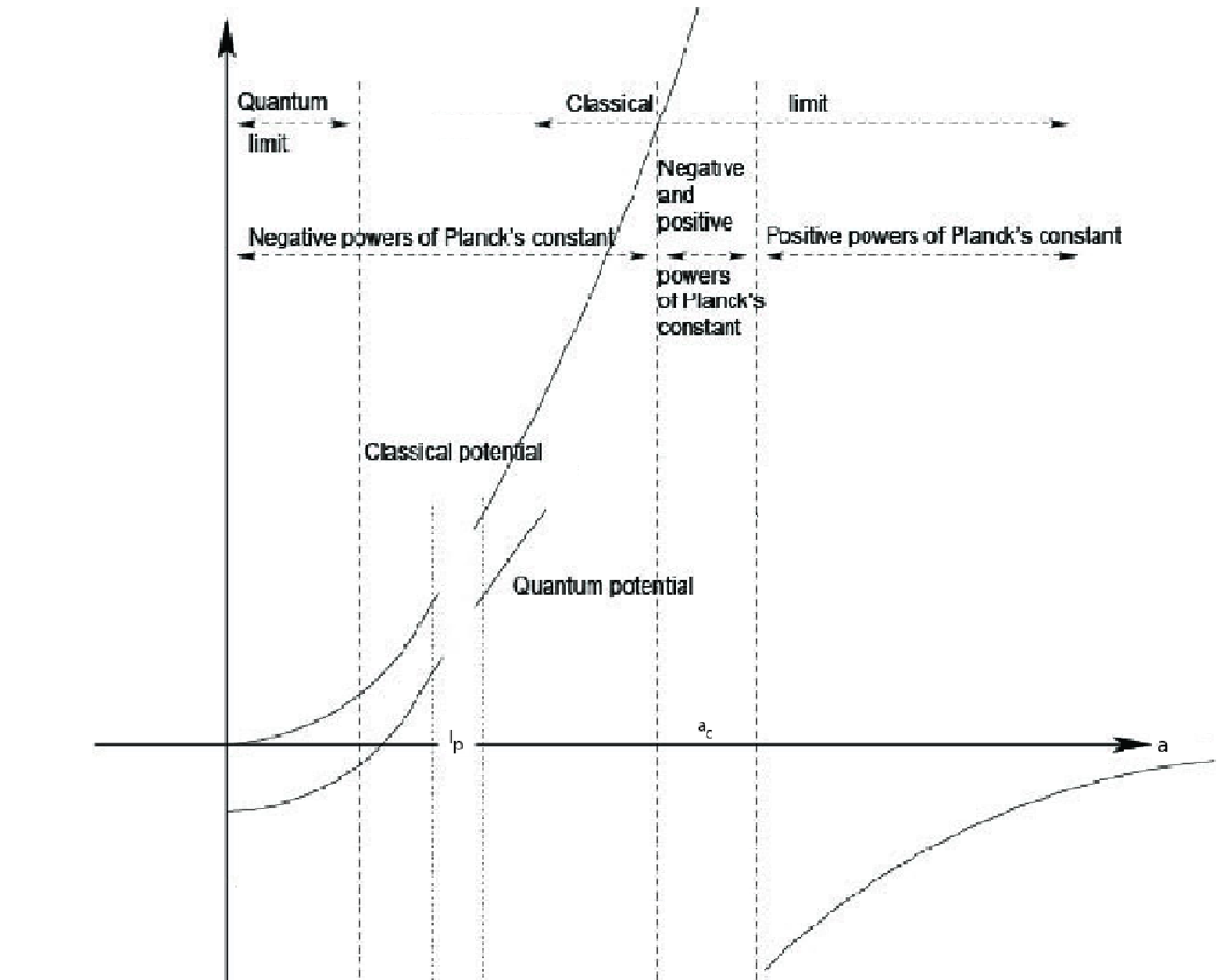}
\end{center}
\caption{Classical and Quantum potentials as a function of 
scale factor.}
\label{f6}
\end{figure}

It is interesting to have more careful look at previous curves. In figures
(\ref{f5}) and (\ref{f6}) a combination of both regimes is plotted. As it is 
clear from these graphs, there are regions in which quantum potential and 
force are less than classical ones, but we have an expansion in terms of 
negative powers of $\hbar$. So it is quiet possible that the criteria 
$\hbar\rightarrow 0$ has nothing to do with the classical limit. The border
of quantum regime is determined by setting the quantum and classical potentials
comparable, which leads to $a\sim a_q=l_p$.

\epsfxsize=5in \epsfysize=5in
\epsfxsize=5in \epsfysize=5in

Although we did anything in a minisuperspace, the result is quiet general. 
Again, here we are interested in gravitational applications, so we start from
the WDW equation which is:
\begin{equation}
\hbar^2G_{ijkl}\frac{\delta^2\psi}{\delta h_{ij}\delta h_{kl}}+V\psi=0
\end{equation}
with the classical potential $V=\sqrt{h}(\ ^{(3)}R-\Lambda)$.
Denoting the couple of indices $ij$ as a single index $A$, simplifies writing:
\begin{equation}
\hbar^2 G_{AB}\frac{\delta ^2\psi}{\delta h_A\delta h_B}+V=0
\end{equation}
The equivalent Bohmian picture is given by:
\begin{equation}
G_{AB}S_{,A}S_{,B}+V+Q=0
\end{equation}
and
\begin{equation}
G_{AB}(R^2S_{,B})_{,A}=0
\end{equation}
where a $ _{,A}$ subscript represents $\delta /\delta h_A$ and the quantum potential is given by:
\begin{equation}
Q=-\frac{\hbar^2R_{,A,B}}{R}
\end{equation}
One can again expand in terms of increasing or decreasing powers of $\hbar$.
\begin{itemize}
\item {\bf Increasing powers of $\hbar$}
Expanding as:
\begin{equation}
S_{,A}=S_{,A}^{(0)}+\alpha S_{,A}^{(1)}+\alpha^2 S_{,A}^{(2)}+\cdots 
\end{equation}
\begin{equation}
R=R^{(0)}+\alpha R^{(1)}+\alpha^2 R^{(2)}+\cdots
\end{equation}
we get order by order:
\begin{itemize}
\item {\bf Zeroth Order:}
In this order we have:
\begin{equation}
G_{AB}S_{,A}^{(0)}S_{,B}^{(0)}+V=0
\end{equation}
\begin{equation}
G_{AB}(R^{(0)2}S_{,B}^{(0)})_{,A}=0
\end{equation}
which is just the classical solution.
\item {\bf First Order:}
In this order we have:
\begin{equation}
2G_{AB}S_{,A}^{(0)}S_{,B}^{(1)}-\alpha G_{AB}\frac{R^{(0)}_{,A,B}}{R^{(0)}}=0
\end{equation}
\begin{equation}
G_{AB}(R^{(0)2}S^{(1)}_{,B}+2R^{(0)}R^{(1)}S^{(0)}_{,B})_{,A}=0
\end{equation}
which leads to first order quantum corrections and so on.
\end{itemize}
\item {\bf Decreasing powers of $\hbar$}
Expanding as:
\begin{equation}
S_{,A}=\alpha^{1/2}(S_{,A}^{(0)}+\alpha^{-1}S_{,A}^{(1)}+\alpha^{-2}S_{,A}^{(2)}+\cdots )
\end{equation}
\begin{equation}
R=R^{(0)}+\alpha^{-1}R^{(1)}+\alpha^{-2}R^{(2)}+\cdots
\end{equation}
we get order by order:
\begin{itemize}
\item {\bf Zeroth Order:}
In this order we have:
\begin{equation}
G_{AB}S^{(0)}_{,A}S^{(0)}_{,B}-G_{AB}\frac{R^{(0)}_{,A,B}}{R^{(0)}}=0
\end{equation}
\begin{equation}
G_{AB}R^{(0)}S^{(0)}_{,A,B}+2G_{AB}R^{(0)}_{,A}S^{(0)}_{,B}=0
\end{equation}
\item {\bf First Order:}
In this order we have:
\begin{equation}
G_{AB}S^{(0)}_{,A}S^{(1)}_{,B}+G_{AB}S^{(1)}_{,A}S^{(1)}_{,B} +G_{AB}\left (
\frac{R^{(1)}R^{(0)}_{,A,B}}{R^{(0)2}}-\frac{R^{(1)}_{,A,B}}{R^{(0)}}\right )+V=0
\end{equation}
\begin{equation}
G_{AB}R^{(0)}S^{(1)}_{,A,B}+G_{AB}R^{(1)}S^{(0)}_{,A,B}+2G_{AB}R^{(0)}_{,A}S^{(1)}_{,B}
+2G_{AB}R^{(1)}_{,A}S^{(0)}_{,B}=0
\end{equation}
and so on.
\end{itemize}
\end{itemize}


\begin{thebibliography}{99}
\bibitem{blas}J. Blaschke, Semi classical approximation beyond the leading order 
in $\hbar$, PhD thesis, 1999.
\bibitem{bal} L.E. Ballentine, Y. Yang and J.P. Zibin, \textit{Phys. Rev. A}, 50, 2854, 
1994;\\
L.E. Ballentine, Quantum mechanics, World Scientific, New York, 1998.
\bibitem{bohm}
L. de-Broglie, \textit{Journ. de Phys.}, 5, 225, 1927;\\ 
L. de-Broglie, \textit{Annales de la Fondation Louis de Broglie}, 12, 4, 1987;\\ 
D. Bohm, \textit{Phys. Rev.}, 85, 2, 166, 1952;\\ 
D. Bohm, \textit{Phys. Rev.}, 85, 2, 180, 1952;\\ 
D. Bohm and B.J. Hiley, The undivided universe, Routledge, London, 1993;\\
D. D\"urr, S. Goldstein, and N. Zangh\`i, \textit{Bohmian Mechanics as the Foundation of 
Quantum Mechanics}, in Bohmian Mechanics and Quantum Theory: An Appraisal, edited by J.T. 
Cushing, A. Fine, and S. Goldstein, Kluwer Academic Publishers, Dordrecht, 1996;\\
R. Tumulka, \textit{Am. J. Phys.} 72, 9, 1220, 2004.
\bibitem{hol} P.R. Holland, The Quantum Theory Of  Motion, Cambridge University Press, 1993.
\bibitem{mos}A. Mostafazadeh, \textit{Nucl. Phys.}, B509, 529, 1998.
\bibitem{allori}V. Allori, and N. Zangh\`i, quant-ph/0112009,  
Based on the talk given at the Biannual IQSA Meeting in Cesena, 
Italy, March 31 - April 5, 2001;\\
V. Allori,D. D\"urr, S. Goldstein and N. Zangh\`i quant-ph/0112005.
\bibitem{foldi} P. Foldi, Highly Non classical Quantum States and Environment Induced 
Decoherence, PhD thesis , quant-ph/0406232.
\bibitem{B1} J.B. Bronzan, \textit{Am. J. Phys.}, 55, 54, 1987.
\bibitem{B2} J.B. Bronzan, \textit{Phys. Rev. A}, 54, 41, 1996.
\bibitem{hor}T. Horiguchi, \textit{Mod. Phy. Lett. A.}, 9, 1429, 1994.
\end{thebibliography}
\end{document}